\documentclass{aa} 
\usepackage{graphicx}
\usepackage{txfonts}
\usepackage{longtable}
\usepackage{subfig}
\usepackage{float}
\usepackage{booktabs}
\usepackage{natbib}
\bibpunct{(}{)}{;}{a}{}{,} 
\usepackage{amssymb}
\usepackage{pifont}
\usepackage{ gensymb }
\usepackage{auto-pst-pdf}
\usepackage{multirow}

\def\mjb{mJy\,beam$^{-1}$}

\def\arcmin{\mbox{$^\prime$}}%
\def\pks{PBC\,J2333.9-2343}

\begin{document}

   \title{Restarting activity in the nucleus of PBC\,J2333.9-2343:}

   \subtitle{an extreme case of jet realignment}

 \author{Hern\'{a}ndez-Garc\'{i}a, L.\inst{1}; Panessa, F.\inst{1}; Giroletti, M.\inst{2}; Ghisellini, G.\inst{3};  Bassani, L.\inst{2};Masetti, N.\inst{2}$^,$\inst{4}; Povi\'c, M.\inst{5}$^,$\inst{6}; Bazzano, A.\inst{1}; Ubertini, P.\inst{1}; Malizia, A.\inst{2}; Chavushyan, V.\inst{7}}

   \institute{INAF - Istituto di Astrofisica e Planetologia Spaziali di Roma (IAPS-INAF), Via del Fosso del Cavaliere 100, 00133 Roma, Italy \\ \email{lorena.hernandez@iaps.inaf.it} \and Istituto di Astrofisica Spaziale e Fisica Cosmica (IASF-INAF), Via P. Gobetti 101, 40129 Bologna, Italy  \and INAF – Osservatorio Astronomico di Brera, via E. Bianchi 46, I–23807 Merate, Italy  \and Departamento de Ciencias F\'{i}sicas, Universidad Andr\'{e}s Bello, Fern\'{a}ndez Concha 700, Las Condes, Santiago, Chile \and Ethiopian Space Science and Technology Institute (ESSTI), Entoto Observatory and Research Center (EORC), Astronomy and Astrophysics Research Division, P.O. Box 33679, Addis Ababa, Ethiopia \and Instituto de Astrof\'{i}sica de Andaluc\'{i}a, CSIC, Glorieta de la Astronom\'{i}a, s/n, 18008 Granada, Spain \and Instituto Nacional de Astrof\'{i}sica, \'{O}ptica y Electr\'{o}nica, Apartado Postal 51-216, 72000 Puebla, M\'{e}xico\\ }

   \date{Received XXX / Accepted XXX}

\authorrunning{Hern\'{a}ndez-Garc\'{i}a et al.}
\titlerunning{PBC\,J2333.9-2343}

 
  \abstract
{\pks\  is a giant radio galaxy which shows different characteristics at different wavebands that are difficult to explain within the actual generic schemes of unification of active galactic nuclei (AGN), thus being a good candidate to host different phases of nuclear activity.}
   {We aim at disentangling the nature of this AGN by using simultaneous multiwavelength data.}
   {We obtained data in 2015 from the Very Long Baseline Array (VLBA), the San Pedro M\'{a}rtir telescope, and the \emph{XMM--Newton} observatories. This allows the study of the nuclear parts of the galaxy through its morphology and spectra, as well as the analysis of the spectral energy distribution (SED). We also reanalysed optical data from the San Pedro M\'{a}rtir telescope from 2009 previously presented in \cite{parisi2012} for a homogeneous comparison.}
   {At X-ray frequencies, the source is unabsorbed. The optical spectra are of a type 1.9 AGN, both in 2009 and 2015, although showing a broader component in 2015. The \emph{VLBA} radio images show an inverted spectrum with self-absorbed, optically thick compact core ($\alpha_c = 0.40 $, where $S_\nu \propto \nu^{+\alpha}$) and steep spectrum, optically thin jet ($\alpha_{j,8-15}=-0.5$). The SED resembles that of typical blazars and is best represented by an external Compton (EC) model with a viewing angle of $\sim$ 3--6 degrees. The apparent size of the large scale structure of \pks\  must correspond to an intrinsic deprojected value of $\sim 7$ Mpc for $\theta_v<10^\circ$, and to $> 13$ Mpc for $\theta_v<5^\circ$, a value much larger than the bigger giant radio galaxy known (4.5 Mpc).}
   {The above arguments suggest that \pks\  has undergone through a new episode of nuclear activity and that the direction of the new jet has changed, in the plane of the sky and is now pointing towards us, making this source from being a radio galaxy to become a blazar, a very exceptional case of restarting activity.}


   \maketitle
%

\newcommand{\xmark}{\ding{55}}%
\newcommand{\cmark}{\ding{51}}%

\section{\label{intro}Introduction}

\begin{table*}
\begin{center}
\caption{\label{observations}Observations of PBC\,J2333.9-2343 studied in this work.}
\begin{tabular}{ccccccc} \hline \hline
Energy band & Instrument  & Date & Exposure (ksec) \\ 
(1) & (2) & (3) & (4)  \\ \hline
X-rays/UV/Optical & \emph{XMM--Newton} (ObsID. 0760990101) & 2015-05-15 & 23 \\
X-rays/UV/Optical & \emph{XMM--Newton} (ObsID. 0760990201) & 2015-11-17 & 25 \\
Optical spectroscopy  & San Pedro M\'artir &  2009-09-18 & 3.6 \\
Optical spectroscopy  & San Pedro M\'artir &  2015-11-07 & 5.4 \\
Radio (8.4, 15, 24 GHz) & \emph{VLBA} &   2015-11-16 & 1.2, 2.0 , 2.7 \\
\hline                      
\end{tabular} 
\caption*{ {\bf Notes.} (Col. 1) observed energies, (Col. 2) instrument or telescope, (Col. 3) date, and (Col. 4)  exposure time.}
\end{center}
\end{table*}

There are billions of galaxies in the Universe, all thought to host a supermassive black hole (SMBH) in their centers \citep[e.g.,][]{salpeter1964,rees1984}. Among them, some show nuclear activity and are known as active galactic nuclei (AGN). In these cases, the SMBH is fed by an accretion disk and about 10\% of the AGN population shows strong biconical relativistic jets \citep[see][and references therein]{kellermann1989,panessa2016}. 

Nowadays, it is believed that all galaxies pass through different phases of nuclear activity. Models predict that AGN activity is a recurrent phenomenon with all galaxies going through active states many times during their lives \citep{hatziminaoglou2001}. 
It has been suggested that evolutionary processes could cause the accretion flow to vanish, giving place to a dormant nucleus -- this is reinforced by the finding of massive black holes in normal galaxies \citep{kormendy1995} -- whereas the activity can be triggered by processes such as interaction and merging of galaxies \citep{bahcall1997} or by secular processes \citep{treister2012}, causing the awaken of the AGN. However, the timescale of these transitions are probably too short compared to the lifetime of the galaxy, and much longer than human-timescales, to be able to observe many sources within these transitional states \citep[e.g.,][]{reynolds1997}.

Nevertheless, many examples of restarted activity in the nuclei of galaxies have been reported in the literature \citep{saikia2009}. Giant radio galaxies \citep[GRG,][]{ishwara1999}, i.e., radio galaxies with lobes of linear sizes larger than 0.7 Mpc, are perfect laboratories to study intermittent activity and galaxy evolution, as they hold 
huge reservoirs of energy that can have spectral ages as old as $\sim 10^7-10^8$ yr -- 
and that can be observed even when not fed by the jets \citep{alexander1987,liu1992}. 
These old lobes represent an historical record of past activity. As the timescale for restarting activity ranges between $\sim 10^4-10^8$ yr \citep{reynolds1997,saikia2010}, in the same large scale dataset we can observe different phases of nuclear activity from the same nucleus. 

Signatures of restarting activity in the jets are observed in GRGs \citep{saikia2009}, in the form of double-double radio galaxies \citep[DDRGs,][]{lara1999,schoenmakers2000,venturi2004,marecki2006,nagai2016}, or as X-shaped radio galaxies \citep[XRGs,][and references therein]{rottmann2001,gopal2012}. Although it is not well established if the formation of DDRGs and XRGs is due to the same mechanism, it is widely accepted that the old extended radio emission is the relic of past radio activity, whereas new jet activity is the one related to smaller scale structures. Different models have been used to explain the 
dichotomy of these structures, such as: the double-AGN model, where two AGN emit two pairs of jets \citep{lal2007}; the back-flow diversion models, where there is a backflow from the active lobes into the wings \citep{leahy1984,kraft2005}; and jet reorientation models, which include precession or other realigning mechanisms \citep{rees1978,dennett2002,merritt2002,liu2004,gopal2012}. 
There are also sources in which restarted activity is deduced from indirect arguments \citep[see e.g.,][]{schoenmakers1998,giovannini2007,konar2008}.

In the present work we focus on the nucleus of PBC\,J2333.9-2343 (also namely PKS\,2331-240, z=0.0475 \citealt{parisi2012}), which is located at RA=23h33m55.2s, DEC=-23d43m41s. This AGN called our attention because it is a GRG \citep{bassani2016} that shows different and incompatible classifications when observed at different frequencies, therefore being a good candidate to host different phases of AGN activity.
Its optical spectrum has been classified as a type 2 Seyfert, i.e., presenting only narrow lines and no broad components \citep[these are typically observed in type 1 AGN,][]{baldwin1981,wilkes1983, parisi2012}. 
Its X-ray spectrum, however, does not show signs of high obscuration, typically seen in Seyfert 2 \citep{risaliti1999,burtscher2016}, and \citet{parisi2012} excluded a Compton-thick nature\footnote{Heavily absorbed sources with column densities greater than $1.5 \times 10^{24}cm^{-2}$ \citep{maiolino1998}.} following classical diagnostic diagrams as in \cite{bassani1999}. It seems indeed that the source behaves as a type 1 AGN at X-rays, i.e., it is unobscured.
Furthermore, at radio frequencies it has many features typical of a blazar and is in fact listed in the
Roma BZCAT as an  object of unknown type \citep{massaro2009} and as a BL Lac by \cite{dabrusco2014} based on WISE colours; it has a flat radio spectrum \citep{healey2007}, it shows a jet feature in a VLBI 8.4 GHz image, it is variable \citep{ojha2004} and also polarized \citep{ricci2004}.

One explanation for the different classifications of PBC\,J2333.9-2343 might be attributed to variability, since the data used to classify this source were obtained in different dates. The alternative scenario is that the different classifications arise from an intrinsic property of the source. In order to investigate the nature of its nucleus, we performed simultaneous multiwavelength observations, including data at radio frequencies, UV, optical, and at X-rays, as well as an analysis of its simultaneous spectral energy distribution (SED).

The paper is organized as follows. Firstly, we present the multiwavelength data -- and its reduction -- obtained from different telescopes and satellites (Sect. \ref{reduction}), whose results are reported (Sect. \ref{analysis}) separately for X-ray, UV, optical, and radio data, including also the analysis of the SED. Later we put together the data obtained at different frequencies to discuss the possible scenario occurring in the inner parts of \pks\  (Sect. \ref{discusion}). Finally, we summarize the main findings obtained from the present work (Sect. \ref{conclusion}).


\section{\label{reduction}Data and reduction}

In the following subsections we present the data used for the analysis and their reduction. The log of all the observations is reported in Table \ref{observations}.

\subsection{XMM-Newton}

We used the proprietary data (PI: P. Parisi) of PBC\,J2333.9-2343, which was observed by \emph{XMM--Newton} twice in 2015 (May 15th and November 17th), only separated by six months apart. We used the data of the EPIC pn camera \citep{struder2001}. The data were reduced 
using the Science Analysis Software (SAS\footnote{http://xmm.esa.int/sas/}), version 15.0.0.
First, good-time intervals were selected, i.e., flares due to solar protons were excluded from the event list. The method we used for this purpose maximizes the signal-to-noise ratio (S/N) of the net source spectrum by applying a different constant count rate threshold on the
single-event light curve with a background of E $>$ 10 keV. We extracted the source (from a circular region of 25'') and background (from a circular region of 40'' free of sources and close to the nucleus) spectra with the {\sc evselect} task.  Response Matrix Files (RMFs) were generated using the task {\sc rmfgen}, and the Ancillary Response Files (ARFs) were created using the task {\sc arfgen}. Light curves in the 0.5--10 keV, 0.5--2.0 keV and 2.0 --10.0 keV energy bands of the source and background were extracted using the task {\sc dmextract} with a 1000 s bin. 

The spectra were groupped to have a minimum of 20 counts per bin using the {\sc grppha task}. This allows the use of the $\chi^2$-statistics. It is worth noticing that the pileup fraction is negligible (below 1\% as estimated with {\sc pimms}\footnote{http://cxc.harvard.edu/toolkit/pimms.jsp} using a power law model).

Data from the Optical Monitor \citep[OM,][]{mason2001} onboard \emph{XMM--Newton} were available in the UVW1 (centred at 2675 \AA), U (centred at 3275 \AA), and V (centred at 5468 \AA) filters in the two dates. The aperture photometry was performed selecting circular regions of 5'' centred on the source and of 20'' for background regions using the processing chain {\sc omichain}. 

\begin{figure}
\centering
\includegraphics[width=0.35\textwidth,angle=270]{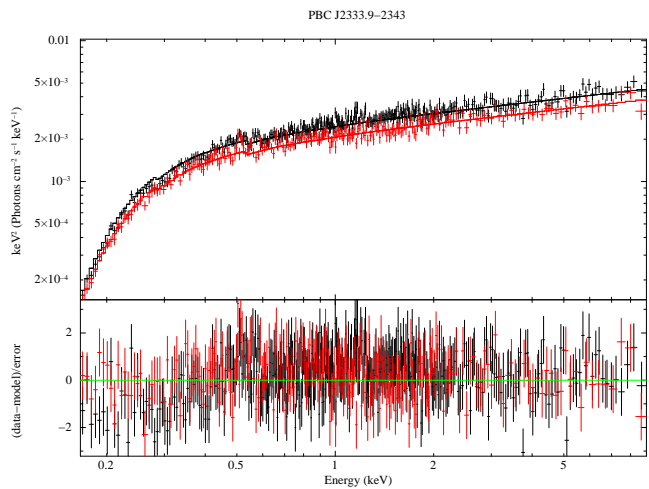}
\caption{\label{spectra} Simultaneous spectral fitting of the \emph{XMM--Newton} spectra of PBC J2333.9-2343 using a single power law model. The black spectrum corresponds to ObsID. 0760990101 and the red spectrum corresponds to ObsID. 0760990201.}
\end{figure}

\begin{table*}
\tiny
\begin{center}
\caption{\label{spectralfit}Spectral parameters of the X-ray spectra of PBC J2333.9-2343 obtained by \emph{XMM--Newton} using a power law model and UV and optical photometry from the OM. }
\begin{tabular}{ccccccc|ccc} \hline \hline
Analysis &  ObsID &  $\Gamma$ & Norm & Flux(2-10 keV) & L(2-10 keV) & $\chi ^2_r$ & UVW1 & U & V\\
              &                           &  & ($\times 10^{-3}$) & ($\times 10^{-11} erg/s/cm^{2}$) &  ($\times 10^{43}$erg/s) & &  & ($\times 10^{-16} erg/s/cm^{2}$) &  \\ 
(1) & (2) & (3) & (4) & (5) & (6) & (7) & (8) & (9) & (10)  \\              \hline
\multirow{2}{*}{Individual} &  0760990101 &  1.77$_{1.75}^{1.78}$ & 2.61$_{2.59}^{2.64}$ & 0.96$_{0.95}^{0.98}$ & 5.07$_{4.98}^{5.16}$ &  1.03 & 8.41$\pm$0.13 & 7.01$\pm$0.09 & 18.78$\pm$0.19 \\ 
 &  0760990201 &  1.77$_{1.76}^{1.79}$ & 2.18$_{2.15}^{2.20}$ & 0.79$_{0.77}^{0.81}$ & 4.16$_{4.07}^{4.27}$ &  0.99 & 8.55$\pm$0.12 & 6.70$\pm$0.07 & 19.55$\pm$0.14 \\ \hline
\multirow{2}{*}{Simultaneous} &  0760990101   & \multirow{2}{*}{1.77$_{1.76}^{1.80}$} & 2.67$_{2.63}^{2.71}$ & 0.97$_{0.93}^{0.99}$ & 5.06$_{4.96}^{5.16}$ & \multirow{2}{*}{1.02} \\
                       &                                      0760990201 &                                            &                                    2.23$_{2.19}^{2.26}$ & 0.78$_{0.75}^{0.80}$ & 4.23$_{4.19}^{4.29}$ & \\
\hline                      
\end{tabular} 
\caption*{ {\bf Notes.} (Col. 1) type of analysis, where individual means that each spectrum was analyzed, and simultaneous means that the two spectra are analyzed together (note that  the normalization of the power law is variable, therefore each of the lines correspond to the values of the spectral parameters of the observations in Col. 2), (Col. 2)  ObsID., (Col. 3) spectral index, (Col. 4) normalization of the power law, (Col. 5 and 6) X-ray intrinsic flux and luminosity in the 2--10 keV energy band, (Col. 7) reduced $\chi^2$, and (Cols. 8, 9, and 10) fluxes from the OM in units of $10^{-16} erg \hspace*{0.1cm} cm^{-2} s^{-1}$. In all cases z=0.0475 and $N_G=1.63 \times 10^{20} cm^{-2}$. The errors are indicated as interval limits and are estimated at the 90\% confidence.}
\end{center}
\end{table*}

\subsection{\label{spm}San Pedro M\'artir telescope}

The nucleus of the galaxy LEDA\,71756, the optical counterpart of PBC\,J2333.9$-$2343, was observed on September 18, 2009, and  on November 7, 2015 with the 2.12 m telescope of the San Pedro M\'artir Observatory (M\'exico), equipped with a Boller \& Chivens spectrograph and a 650$\times$2048 pixels E2V-4240 CCD. 
In this section we report only the data reduction for the spectrum obtained in 2015, since the spectrum from 2009 was reduced and published in \cite{parisi2012} and we refer the reader to this paper for more details on the data reduction. However, a stellar population subtraction has been applied in this work for the first time to both spectra (see details in Sect. \ref{opticalresults}).

Three 1800-s spectra with a dispersion of $\sim$2.3 \AA/pix were acquired starting at 03:15 UT on November 7,  2015; they nominally covered the 3600--8200 \AA~wavelength range. 
The spectra were acquired on the galaxy nucleus with a slit aperture of 2.5 arcsec.
Data reduction was carried out using standard procedures for both bias subtraction and flat-field correction with 
IRAF\footnote{{\tt http://iraf.noao.edu/}}.

The wavelength calibration was carried out with neon-helium-copper-argon lamps; the uncertainty was $\sim$0.5 \AA~according to the position of the background night sky lines. The flux calibration was performed using the catalogued spectroscopic standard star G191$-$B2B \citep{massey1988}.
The three spectra were then stacked together to obtain one final spectrum with the best possible S/N. 

It is worth noticing that the values of the airmass (1.7-1.8) during the observations was high and, consequently, the separation between blue and red wavelengths is large, thus the spectroscopic flux measurements need to be taken with caution. However, the estimates on line widths and ratios are reliable, providing a robust classification of the source.

\subsection{Very Long Baseline Array (VLBA)}

We observed the nucleus with the NRAO\footnote{The National Radio Astronomy Observatory is a facility of the National Science Foundation operated under cooperative agreement by Associated Universities, Inc.} Very Long Baseline Array (VLBA) on 2015 November 16. We observed for two hours at 8.4, 15, and 24 GHz, with a net on source time of 20, 33, and 45 minutes at each frequency, respectively. Ice and cold weather affected the focus/rotation mount at Mauna Kea, so that data were obtained from that station only for a limited part of the observations. No other issues were reported.

The observing frequency band was divided in $16\times 32$ MHz baseband channels, in dual polarization, for a total recording rate of 2 Gbps. We carried out a standard calibration in AIPS, following the new prescription for amplitude calibration described by \citet{Walker2014}. We removed single band delays with pulse-cal table information, and the residual delays and phase rates by fringe fitting the target source itself. We then averaged the frequency channels and produced the final images in Difmap, with several cycles of phase-only and phase-and-amplitude self calibration, with solution intervals of decreasing length. The noise levels are within a factor $2$ of the predicted thermal values, which is reasonable given the non-ideal coverage of the $(u,v)$-plane for short observations of a low declination source, and the rather complex structure of the object.

\section{\label{analysis}Data analysis and results}

\subsection{\label{xrayresults}X-ray spectroscopy and UV/optical photometry}

Firstly, we checked whether X-ray variations are observed on short timescales (i.e., intra-day variability). We analyzed the \emph{XMM--Newton} light curves in three energy bands, the soft (0.5--2 keV), hard (2--10 keV), and total (0.5--10 keV). We calculated the normalised excess variance, $\sigma^ 2_{NXS}$, following prescriptions in \cite{vaughan2003} \citep[see also][]{omairavaughan2012}. 
None of the light curves showed short-term variations, i.e., all are compatible with zero within the errors.

Later, we performed the spectral analysis using XSPEC v. 12.9.0. We assume a redshift of z=0.0475 \citep{parisi2012} and a Galactic absorption of $N_{Gal}= 1.63 \times 10^{20} cm^{-2}$ as obtained within {\sc ftools} \citep{dickey1990,kalberla2005}.
Different models were tested in order to select the simplest one which fits well the X-ray data. Firstly, these models were fitted to each spectrum individually. Initially, we fitted a single power law representing the AGN continuum. This model represented well the data, but we tried to add different components to check whether there was an improvement of the fit. 
We added neutral absorption using a {\sc zwabs} component, then changed the absorption by a partial covering component ({\sc pcfabs} in XSPEC), added a thermal component ({\sc mekal} in XSPEC) or another power law, and also fitted these components absorbed by an intrinsic column density.
The $\chi^2/d.o.f=\chi^2_r$ and $F$-$test$ were used to select the best-fit model, considering an improvement on the spectral fit when $\chi^2/d.o.f$ is as closest as possible to the unity and the $F$-$test$ results in a value larger than $10^{-5}$. None of these components improved the fits, thus the best representation of the data is obtained by a single power law model. The results of the spectral parameters and X-ray fluxes and luminosities are reported in Table \ref{spectralfit} (marked as $Individual$ in Col. 1).
The individual spectral fits agree well with that reported by \cite{parisi2012} for \emph{Swift} data.

As a second step, we fitted the two \emph{XMM--Newton} spectra simultaneously. This process allows to search for X-ray spectral variability \citep{lore2013}. We found that linking all the parameters to each other does not result in a good spectral fit  ($\chi^2/d.o.f.=1865.9/1291$).

We therefore let the spectral index, $\Gamma$, and the normalization of the power law, $Norm$, vary in the model one-by-one. The model where $Norm$ varied showed an improvement ($F$-$test=1 \times 10^{-100}$), whereas varying $\Gamma$ did not show an improvement of the fit.
We also let the two spectral parameters vary together (i.e., $\Gamma+Norm$), but no improvement of the spectral fit was detected.

Thus, our results show that PBC\,J2333.9-2343 has undergone a small luminosity decrease (16\%) of its nuclear power during the timescale of six months.
PBC\,J2333.9-2343 has an intrinsic column density compatible with the Galactic one (we estimated an upper limit of $N_H < 2.7 \times 10^{20} cm^{-2}$ on the intrinsic absorption), a spectral slope of about 1.8, and a variable normalization of the power law.
In Fig. \ref{spectra} we present the simultaneous spectral fit for \emph{XMM--Newton} data when varying $Norm$. The results of the spectral fit can be found in Table \ref{spectralfit}  (marked as $Simultaneous$ in Col. 1). 
It is worth noting the presence of residuals in the 0.5-1 keV energy range that might be due to emission/absorption features. We added Gaussian components to represent these lines in the spectral fitting but none of them were significant for the spectral fit (using C-statistics). We also checked the data from the reflection grating spectrometer (RGS), which show hints of spectral lines, but a longer observation is needed in order to reproduce these spectral features.

Finally, we obtained the UV and optical photometry from the OM in three filters; UVW1, U, and V, whose measurements are presented in  the right part of Table \ref{spectralfit}. Variations between the two dates (separated by six months) observed by OM were not detected in the UVW1 filter, since the fluxes are compatible within the errors. A small variation of 4\% was detected both in the optical U and V filters.

\begin{figure*}
\centering
\includegraphics[width=0.9\textwidth,trim=0cm 0cm 0cm 1.1cm,clip]{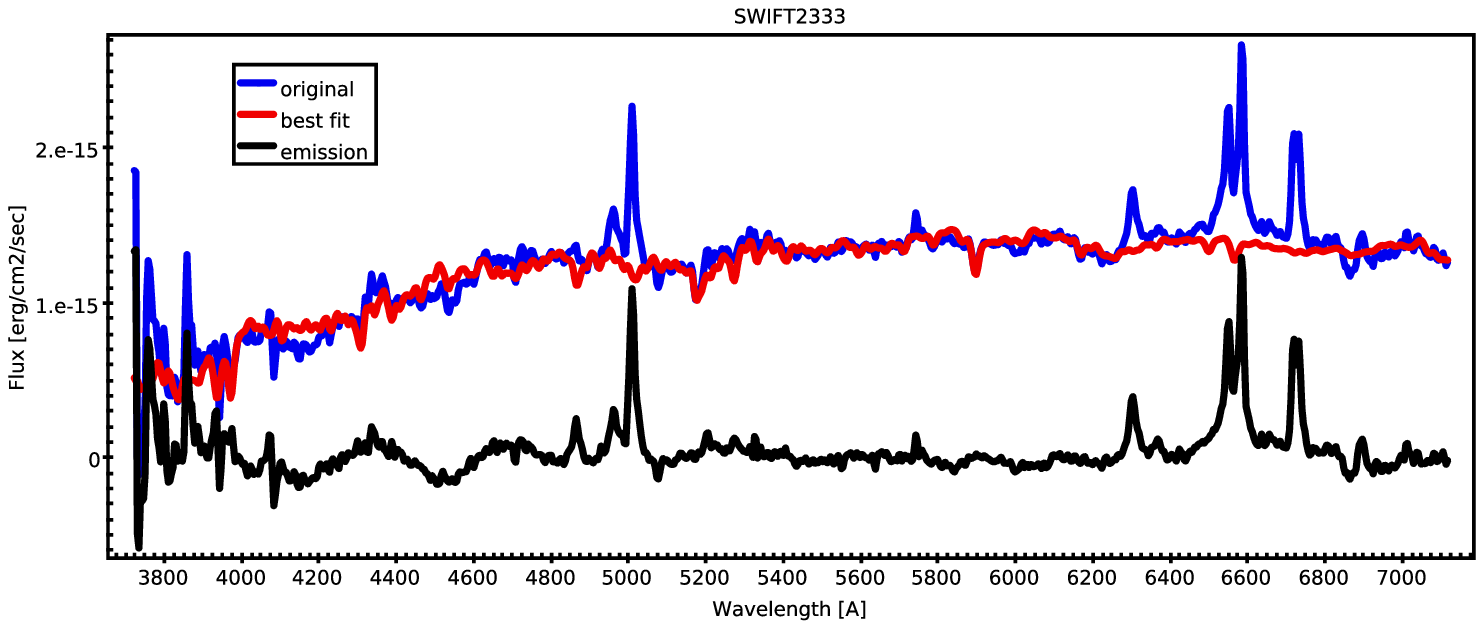}

\vspace*{-0.5cm}

\includegraphics[width=0.9\textwidth,trim=0cm 0cm 0cm 1.1cm,clip]{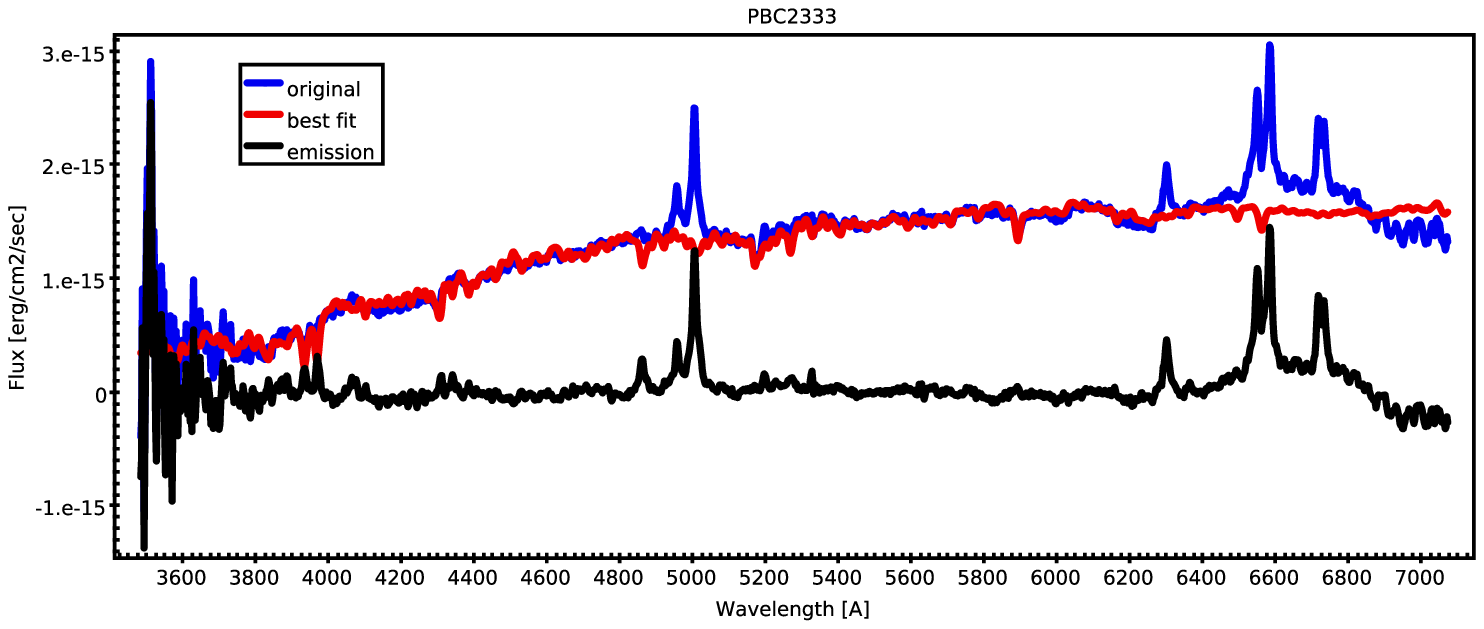}
\caption{\label{optspectra} Optical spectra of PBC J2333.9-2343 obtained with the San Pedro M\'artir Observatory in 2009 (up) and 2015 (bottom). The blue corresponds to the original spectrum, the red to the best-fit model, and the black to the stellar population subtracted spectrum (i.e., observed - best-fit model).}
\end{figure*}

\begin{table}
\caption[]{\label{opticalprop}Main properties of the optical spectra (see Fig. \ref{optspectra}).}
\begin{center}
\begin{tabular}{lcc} \hline \hline
Property & 2009 & 2015 \\ 
(1) & (2) & (3) \\ \hline
S/N & 54.6$\pm$1.1 & 57.1$\pm$1.1  \\
 log$M_{\bigstar}$ &  10.77$\pm$0.18  &  10.61$\pm$0.19   \\
 log$M_{\bigstar} (initial)$ &  11.07$\pm$0.22  & 10.87$\pm$0.23  \\
log(\={Age}) & 9.78$\pm$0.12  & 9.52$\pm$0.12  \\
SFR & 6.55$\pm$1.44  & 4.16$\pm$0.91  \\
Flux($H_{\beta}$) & 5.61$\pm$0.67 & 5.56$\pm$1.13 \\

 Flux([O$_{\sc III} (4959)$]) & 5.52$\pm$1.02 & 8.23$\pm$1.13 \\

 Flux([O$_{\sc III} (5007)$]) & 19.23$\pm$1.02 & 22.10$\pm$1.13 \\

 Flux([O$_{\sc I} (6300)$]) & 9.98$\pm$1.14 & 9.30$\pm$0.85 \\

 Flux([N$_{\sc II} (6548)$]) & 13.10$\pm$9.73 & 13.49$\pm$1.01 \\
            
Flux($H_{\alpha}$) & 3.12$\pm$0.97 & 5.66$\pm$1.01 \\
            
 Flux([N$_{\sc II} (6584)$]) & 19.37$\pm$0.97 & 20.00$\pm$1.01 \\
                                  
 Flux([S$_{\sc II} (6717)$]) & 11.15$\pm$0.88 & 8.00$\pm$0.58 \\
           
 Flux([S$_{\sc II} (6731)$]) & 8.87$\pm$0.88 & 7.47$\pm$0.58 \\ \hline
\end{tabular}
\caption*{ {\bf Notes.} (Col. 1) Property, where all the fluxes are measured in units of $10^{-15} erg \hspace*{0.1cm} s^{-1}$, (Col. 2) value for the spectrum taken in 2009, and (Col. 3) value for the spectrum taken in 2015 . For more information about flux measurements and their reliability see Sect. \ref{spm}.}
\end{center}
\end{table}

\subsection{ \label{opticalresults} Optical spectroscopy}

We estimated the stellar population using the STARLIGHT V.04 stellar population synthesis code \citep{cid2005,cid2009} for the spectra obtained in 2009 and 2015.  This code decomposes an observed spectrum in terms of a superposition of a base of simple stellar populations of various ages and metallicities, outputting physical properties as, e.g., the star-formation and chemical enrichment histories of a galaxy.
Before the subtraction of the stellar population, the spectra were corrected for Galactic extinction, K-corrected, and moved to rest-frame. We used the fits based on the templates from \cite{bruzual2003}, with solar metallicity, and 25 stellar ages in the range [$1 \times 10^{6} - 1.8 \times 10^{10}$] yr, and the extinction law of \cite{cardelli1989}. During the fits, all the areas with emission lines, noise, and atmospheric absorption were masked.
It is worth noticing that both spectra were affected by detector fringing above 7000 \AA, and thus they were cut after the $[S${\sc ii}$]$ line at about that wavelength. Finally, the 3600--7000 \AA\ wavelength range was used for the analysis. 
We caution that the CCD sensitivity in the blue region of the 2009 spectrum is low, resulting in strong residuals in the 4000-4800 \AA\  wavelength range when performing the stellar population subtraction as it cannot be well fitted with STARLIGHT. 
The spectra subtracted by the stellar population from the STARLIGHT fits are presented in Fig. \ref{optspectra}. 

The main properties of the optical spectra are presented in Table \ref{opticalprop}. The S/N was obtained directly from the best-fit models. To measure the current ($M_{\bigstar}$) and initial ($M_{\bigstar}(initial)$) mass in stars, star formation rate (SFR), and mean age (log(\={Age})) we followed the same procedures as in \citet[][and references therein]{povic2016}. The independent measurements of $M_{\bigstar}$, $M_{\bigstar}(initial)$, log(Age) and SFR as listed in Table \ref{opticalprop} agree well within the errors. 
Line fluxes were obtained by means of splot IRAF task by fitting a single gaussian function, while the errors were measured taking into account the rms of the continuum. 
The resulting best-fit models show that both spectra are completely dominated by old stellar populations with ages $>10^9$ yr. 

The optical spectrum of PBC J2333.9$-4$2343 obtained in 2015 does not differ substantially from the one obtained in 2009 (first presented in \cite{parisi2012}, without the stellar population subtraction). Indeed, the line fluxes of the emission lines are similar between the two spectra, as well as the continuum level (see Table \ref{opticalprop})\footnote{It is worth noticing here that the measurements of $H_{\alpha}$, [N{\sc ii}], and [S{\sc ii}] might be affected by the broad component and thus they cannot be completely reliable.}. We also remark that source continuum and narrow emission lines do not show significant variability among the spectra acquired in the framework of the present paper.

\begin{figure*}
\includegraphics[width=0.87\paperwidth]{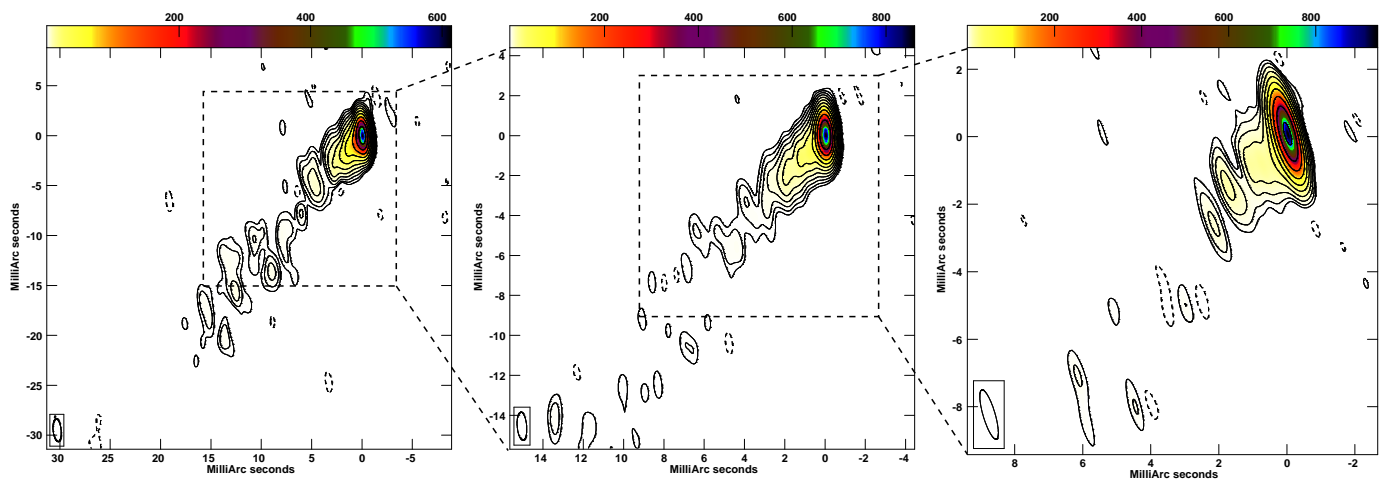}
\caption{VLBA images of \pks\ at 8, 15, and 24 GHz (left to right). Contours are traced at $(-1,1,2,4,\dots)\times$ 0.4, 0.4, 1.6 \mjb\ in each image, respectively. The restoring beam is shown in the bottom left corner of each panel and its value is given in Table~\ref{t.logvlba}. The dashed box in the left panel indicates the region displayed in the middle one; the one in the middle panel indicates that of the right hand side one. The colour scale range shows surface brightness in mJy/beam between 0 (yellow) and the image maximum (blue), as shown in the top wedge of each panel. \label{f.vlba}}
\end{figure*}

\begin{table*}
\caption{VLBA image parameters. \label{t.logvlba} }
\begin{center}
\begin{tabular}{cccccc} \hline \hline
$\nu$ & Duration & HPBW & $\sigma_\nu$ & $I_\nu$ & $S_\nu$ \\
(GHz) & (min) & (mas $\times$ mas, $^\circ$) & (\mjb) & (\mjb) & (Jy) \\
(1) & (2) & (3) & (4) & (5) & (6) \\
\hline
8 & 25 & $2.3\times0.8, 5.4$ & 0.08 & 610 & 0.91 \\
15 & 33 & $1.4 \times 0.5, 4.7$ & 0.08 & 810 & 1.16 \\
24 & 45 & $1.5 \times 0.35, 16.3$ & 0.34 & 940 & 1.23 \\ \hline
\end{tabular}
\caption*{ {\bf Notes.} (Col. 1) Frequency, (Col. 2) duration, (Col.\ 3): half peak beam width (HPBW); (Col.\ 4): image noise; (Col.\ 5): image peak brightness; (Col.\ 6): total cleaned flux density. }
\end{center}
\end{table*}

\begin{figure}
\includegraphics[width=0.4\paperwidth]{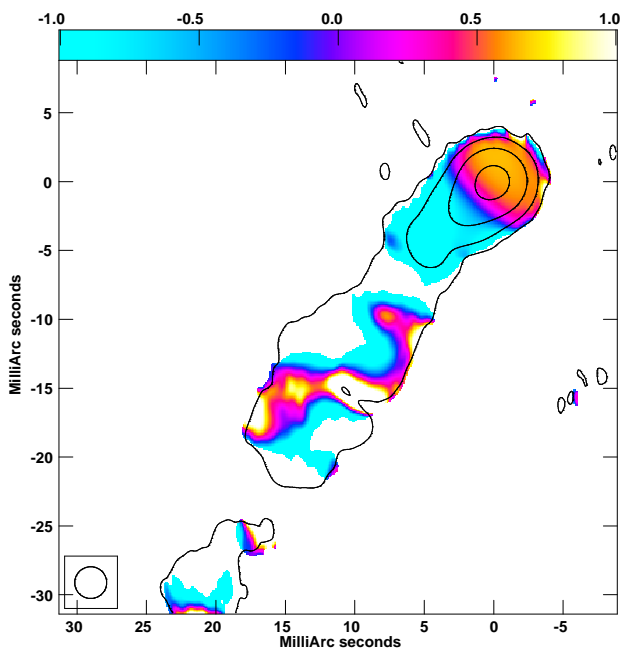}
\caption{VLBA spectral index image of \pks\ between 8 and 15 GHz, overlaid with 8 GHz total intensity contours. The colour scale range is $-1 < \alpha <+1$ and it is shown by the wedge on the top; contours are traced at $(1,10,100,1000)\times$ 0.35 \mjb. \label{f.spix}}
\end{figure}

Nevertheless, one peculiarity is found when comparing our results with those reported in \cite{parisi2012}. In their paper, the authors classified this source as a type 2 Seyfert. In the present analysis, using the same spectrum from 2009 and thanks to the subtraction of the stellar population, we report the existence of a broad component at the foot of the H$_\alpha$+[N{\sc ii}] complex. However, we are not able to measure it in the observation from 2009 because it is very weak. Recent spectra taken in 2016 confirm the current presence of the broad H$_\alpha$ line and will be presented in a forthcoming paper (Hern\'{a}ndez-Garc\'{i}a et al., in prep.).

The broad component in the H$_\alpha$+[N{\sc ii}] complex is more marked in the 2015 spectrum, which has a FWHM of about 370 \AA, corresponding to a velocity of approximately 16700 km s$^{-1}$, and a flux of $ 1.6 \times 10^{-13} erg \hspace*{0.1cm} cm^{-2}s^{-1}$ (although see Sect. \ref{spm}). This might suggest a small amount of variability in the broad line region (BLR) between the two dates, although we cannot quantify it.

Therefore, since a broad component was found in H$_\alpha$ both in 2009 and 2015, but not in H$_\beta$, we suggest that this source is a type 1.9 AGN according to the classification criteria given by \cite{osterbrock1981}.

We also note that the optical spectra show some evidence of a “blue wing” affecting all nebular lines which may trace an ionised outflow. A detailed analysis of this outflow component will be presented in a forthcoming paper (Hern\'{a}ndez-Garc\'{i}a et al., in preparation).

Finally, we can estimate the black hole mass, $M_{BH}$, using the relation in \cite{greene2005}, which relates the FWHM($H_\alpha$) and the flux of the $H_\alpha$ broad component\footnote{Note that the measurement of the flux has to be taken with caution (see Sect. \ref{spm}).} with $M_{BH}$. From this approach and using the data from 2015, we obtained $M_{BH} = 5.7 \pm 2.3 \times 10^8 M_{\odot}$. Alternatively, we can estimate the $M_{BH}$ from the velocity dispersion derived with STARLIGHT. Assuming an average instrumental resolution (from the FWHM of the sky lines) of 9\AA\ in the range between 4000-7000\AA (i.e., a central wavelength of 5500\AA), this yields a corrected velocity dispersion of 330 km/s. Using the M-$\sigma$ relation in \cite{gultekin2009} we estimate a black hole mass of  $M_{BH} = 5.6 \pm 1.3 \times 10^8 M_{\odot}$.

\subsection{\label{radioresults}Radio}

The source is well detected at all three frequencies (8, 15, and 24 GHz, see Table \ref{t.logvlba} for details on the image parameters at each frequency). It is characterized by a compact bright core and a jet (see Fig.~\ref{f.vlba}). The overall source flux is fairly similar at the various frequencies, indicating a flat radio spectrum.
In Fig.~\ref{f.spix}, we show the spectral index image between 8 and 15 GHz, obtained from images produced using the same baseline length range ($8-240 \, {\rm M}\lambda$) and restoring beam (circular, 2.3 mas HPBW). The spectral index $\alpha$ is defined such that $S(\nu)\sim\nu^{+\alpha}$.

The core is the main feature in the VLBA images. Its peak brightness varies between 610 and 940 \mjb, increasing with frequency.  The mean spectral index is $\alpha_{\rm c} = 0.40\pm0.04$ based on a linear fit to these values, which is in agreement with what shown by the spectral index image (Fig.~\ref{f.spix}). This behaviour indicates that the core region is in an optically thick regime.
The brightness temperature is about $10^{12}$ K, providing a first indication that the Doppler factor $\delta = [ \Gamma (1-\beta\cos\theta_v)]^{-1}$ is greater than unity; with standard notation, $\Gamma=(1-\beta^2)^{-1/2}$, $\beta=v/c$, $v$ is the jet velocity and $\theta$ is the jet viewing angle.

The core is at the base of a one-sided jet, which extends in position angle $\varphi = 133^\circ$. The jet is brighter at 8 GHz, where it extends out to over 60 mas ($\sim 53$ pc), as best seen in naturally weighted images. The brighter jet at 8 GHz is in agreement with optically thin synchrotron emission, as indicated also by the steep values of the spectral index seen in Fig.~\ref{f.spix}; the average value in the jet region is $\alpha_{\rm j, 8-15}=-0.5$. 

The large scale structure of \pks\ is two-sided, indicating that asymmetries on parsec scales are most likely the result of Doppler beaming. We set a lower limit to the jet/counter-jet brightness ratio $R$ using the jet brightness at $\sim$ one beam from the core and the $1\sigma$ rms noise level on the counter-jet side; in particular, for the 8 GHz image, we get $R>10^3$. In turn, by using the observed spectral index value of $\alpha_{\rm j}$, we get $\beta \cos \theta_v > 0.76$. Interestingly, this immediately implies that $\theta_v < 40^\circ$. The comparison of the jet position angle (p.a.) on parsec and kiloparsec scales does not suggest the presence of strong viewing angle changes; indeed, \citet{Graham2014} reported a jet p.a.\ of $\varphi=139^\circ$ and the lack of p.a.\ changes in the parsec scale of \pks, based on the images in \citet{Fomalont2000}, \citet{ojha2004}, and in the radio fundamental catalogue (RFC, version rfc2013d).

Finally, we can compare our image at 8 GHz with those at the same frequency reported by \citet{ojha2004} and in the RFC. We find flux density variability, with the source being $\sim300$ mJy brighter in the present epoch. The jet p.a. has not changed significantly in comparison to the RFC; also the difference to the p.a.\ reported by \citet[][$\varphi=121^\circ$]{ojha2004}, is not highly significant (in particular if we consider that the array used by these authors had a very limited coverage of the $(u,v)$-plane). The time separation between the three dataset is more than 11 years, making the component identification highly uncertain. At the redshift of the source, an apparent proper motion with velocity $\beta=1$ would correspond over this time range to a displacement of $\sim4.2$ mas. This means that, if they are travelling with $\beta=1$, all the components present in the archival data have now moved to the region where the jet is transversally resolved and the compact component have been disrupted.

\subsection{\label{sed}Spectral energy distribution (SED)}

The analysis of the spectral energy distribution (SED) was carried out using the simultaneous data presented in this work.

The strong and flat radio emission, the large radio--loudness and
the strong X--ray emission, all suggest that the flux
produced by the jet is a very important component
for modelling the entire SED, that resembles closely the
SED of typical blazars \citep{ghisellini2015}.

To the aim of reproducing the overall SED, we use the model
fully described in \cite{ghisellini2009}.
It is a one--zone leptonic model that assumes that the
emitting region is a sphere of radius $R=\psi R_{\rm diss}$,
where $R_{\rm diss}$ is the distance from 
the black hole, and $\psi$ is the jet opening angle, assumed to be
conical.
The source is moving at a relativistic velocity $\beta_c$
corresponding to the bulk Lorentz factor $\Gamma$.
This region is observed at a viewing angle $\theta_v$,
so that the relativistic Doppler factor is 
$\delta\equiv 1/[\Gamma(1-\beta\cos\theta_v]$.

The energy distribution of the emitting particles is derived
solving a continuity equation assuming that particles  
are uniformly injected throughout the source, and considering
radiative cooling, electron--positron pair production
and the contribution of these pairs to the observed flux.
The considered emission processes are synchrotron, synchrotron self--Compton (SSC)
and inverse Compton scattering off photons produced externally to the jet (EC),
by the disk, by the BLR and by the molecular torus.

The emitting source is compact, as in other blazars. 
In our case, $R_{\rm diss}=$300--400 Schwarzschild radii.
Such a compact region produces a synchrotron spectrum that is self--absorbed
below $\sim$100 GHz, and therefore cannot account for the radio flux at smaller frequencies.
This has to be produced by other, more extended, portions of the jet. 

\begin{table*} 
\caption{Adopted parameters for the jet models.}
\centering
\begin{tabular}{l l l l l l l l l l l l l l l l l l}
\hline
\hline
&$M$ &$L_{\rm d}$ &$L_{\rm d}/L_{\rm Edd}$ &$R_{\rm diss}$ &$R_{\rm BLR}$ &$R_{\rm torus}$ 
&$P^\prime_{\rm e, jet, 45}$  &$B$ &$\Gamma$ &$\theta_v$  \\ 
~(1) &(2) &(3) &(4) &(5) &(6) &(7) &(8) &(9) &(10) &(11) \\
\hline   
EC  &3.0e9  &0.047 &1.3e--4 &360  &22  &540  &1e--3 &0.14  &10 &3        \\   

SSC &---  &---   &---     &90   &--- &---  &7e--3 &0.17  &13 &6      \\  
\hline
\hline 
\end{tabular}
\vskip 0.4 true cm
\caption*{{\bf Notes.} 
Col. (1) model type
Col. (2) black hole mass in solar masses;
Col. (3) disk luminosity in units of $10^{45}$ erg s$^{-1}$;
Col. (4) disk luminosity in units of the Eddington luminosity;
Col. (5) distance of the dissipation region from the black hole, in units of $10^{15}$ cm;
Col. (6) size of the BLR, in units of $10^{15}$ cm;
Col. (7) size of the torus, in units of $10^{15}$ cm;
Col. (8) power injected in the jet in relativistic electrons, calculated in the comoving 
frame, in units of $10^{45}$ erg s$^{-1}$;
Col. (9) magnetic field in G;  
Col. (10) bulk Lorentz factor;
Col. (11) viewing angle in degrees;
}
\label{para}
\end{table*}

\begin{figure} 
\vskip -0.6 cm
\includegraphics[width=8.5cm,height=9cm]{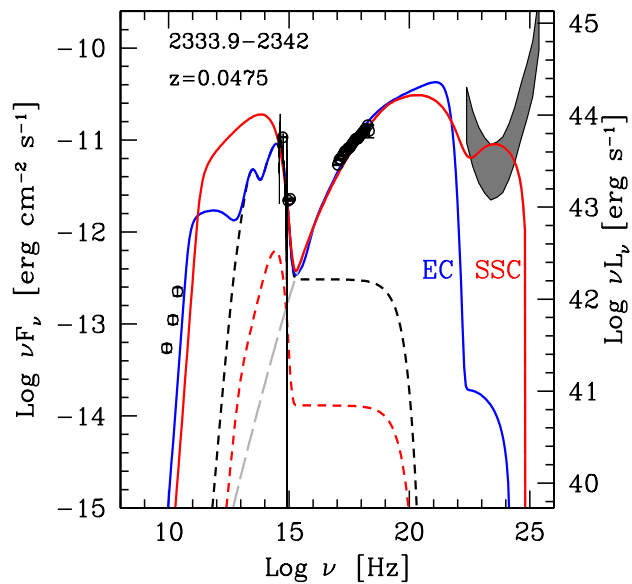} 
\vskip -0.5 cm
\caption{
The SED of PBC J2333.9--02343 using simultaneous data from the \emph{VLBA} and \emph{XMM--Newton} observatories.
The grey stripe corresponds to the 5$\sigma$ sensitivity of
{\it Fermi}/LAT after 1 year (top margin) and 5 years (bottom margin)
of observations.
The solid lines refer to the theoretical models:
for the continuous red line it is assumed that the seed photons 
used for the Inverse Compton scattering are the synchrotron
ones produced by the same electrons (SSC);
for the continuous blue line the process is External Compton (EC)
with seed photons produced by the torus.
The upturn of the SSC spectrum in the $\gamma$--ray band is due to the
second order Compton scattering, and conflicts with the
fact that this source is not detected by {\it Fermi}/LAT.
The black dashed line represents the disk; the red dashed line represents the corona.
}
\label{sedfig}
\end{figure}

Fig. \ref{sedfig} shows the overall SED of PBC\,J2333.9-2343 together with 
our models. The grey area represents the 5$\sigma$ sensitivity of
{\it Fermi}/LAT after 1 year (top margin) and 5 years (bottom margin)
of observations.
Table \ref{para} lists the parameters used for the EC model (blue line) and the 
SSC one (red line). We recall that details on the models and on how to estimate the parameters can be found in  \cite{ghisellini2009}.

We have searched in the Fermi archive\footnote{http://fermi.gsfc.nasa.gov/ssc/} for indications of $\gamma$-ray emission from the source but have not detected any, as reported also by \cite{giommi2012}.
Using the Fermi LAT photon event and spacecraft data query and online data analysis tool\footnote{http://tools.asdc.asi.it/?\&searchtype=fermi}, we find that \pks\  is not close to the Galactic plane nor to a blazar, thus the non detection cannot be related to these problems. Moreover, we estimated the Fermi upper limit for this source, which lies within the grey area in Fig. \ref{sedfig}.
Since the detection of this source is expected under the SSC scenario, we suggest that the preferred main emission process in \pks\  is EC. 

From the EC model, and assuming a standard accretion disc \citep{shakura1973}, we estimate a $M_{\rm BH}=3 \times 10^9 M_\odot$ from the frequency of 
the peak of the accretion disc spectrum and its flux. However, the peak is not well determined, therefore the uncertainty is about a factor two. It is worth noting that this value is roughly consistent with that obtained from the optical data (see Sect. \ref{opticalresults}), taking into account that both measurements are not accurate.
The accretion disk emits a luminosity $L_{\rm d}\sim 5\times 10^{43}$ erg s$^{-1}$,
therefore the Eddington ratio is $R_{Edd} = L_{\rm d}/L_{\rm Edd}\sim 1.3\times 10^{-4}$. 
For the modelling, we use a standard disk spectrum \citep{shakura1973},
even if the Eddington ratio indicates that theoretically we are likely in the advection dominated accretion flow \citep[ADAF,][]{narayan1994} regime.
In any case, the seed photons that we use for the inverse Compton process are
produced by the torus, not by the BLR clouds.
If the disk is emitting the same luminosity, but in ADAF mode, we would have a
different optical--UV spectrum. 
However, the energy density of the radiation
produced by the torus would remain unchanged (since the torus reprocesses 
all the incoming radiation, largely independently from its spectrum). 

To our aim, the important and robust result is that the observed level 
of radio emission and X--ray flux and slope can be reproduced only if the jet is observed
at small (3--6 degrees) viewing angles.
At larger angles the SED cannot be well reproduced. In particular, the Doppler factor (hence the flux enhancement) decreases so much to require 
impossibly large intrinsic (i.e. comoving) luminosity to account for the observed data.
In addition, larger angles imply an increasing importance of the second order Compton scattering, that contributes in the gamma-ray band with a flux that should become visible by Fermi/LAT. Nevertheless, high energy data (\emph{Swift}/BAT, \emph{NuSTAR}, \emph{INTEGRAL}, \emph{Fermi}) would be useful to better constrain our model parameters, in particular $\theta_v$.

\begin{figure*}
\includegraphics[width=0.4\paperwidth]{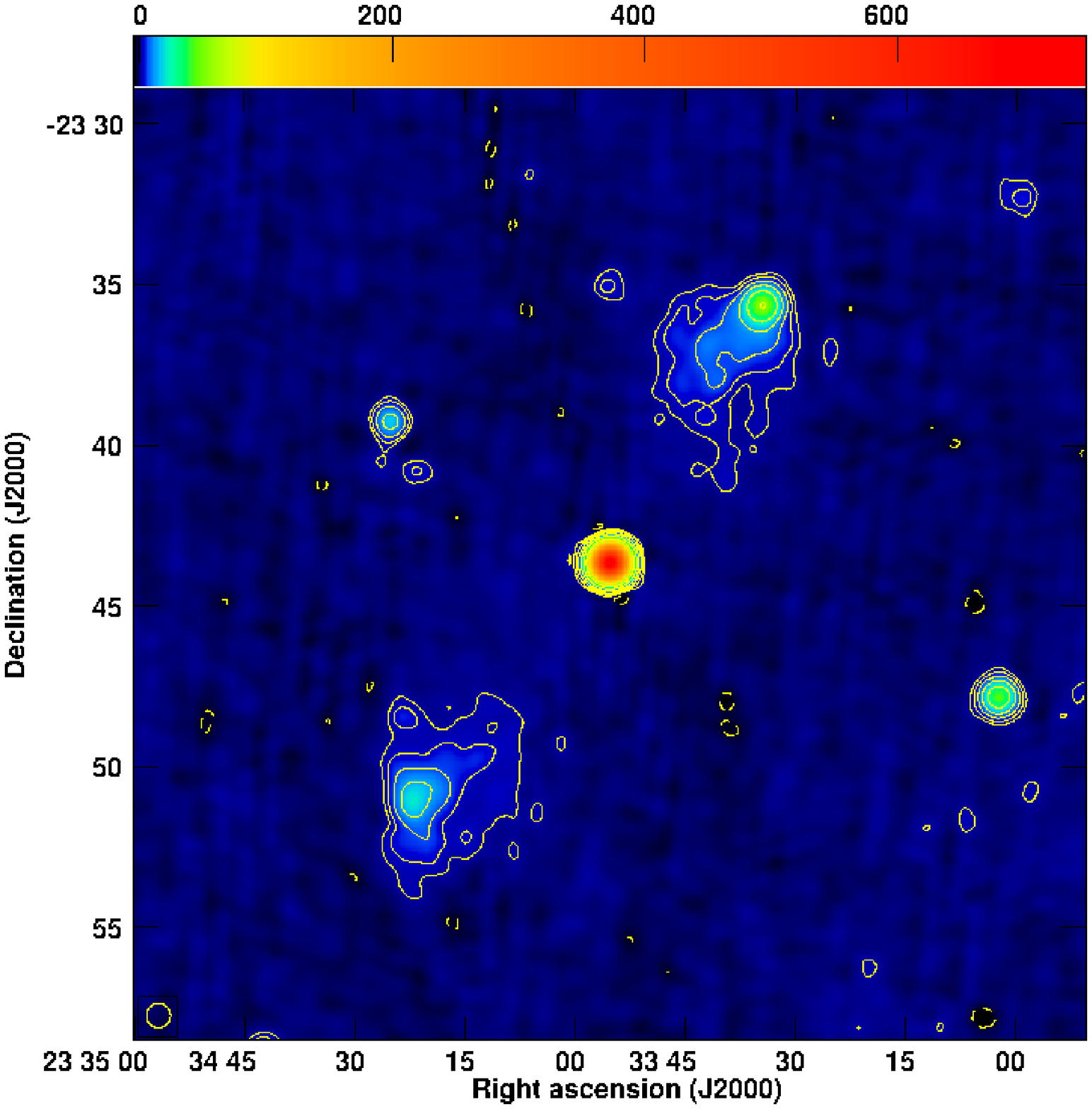} 
\includegraphics[width=0.45\paperwidth]{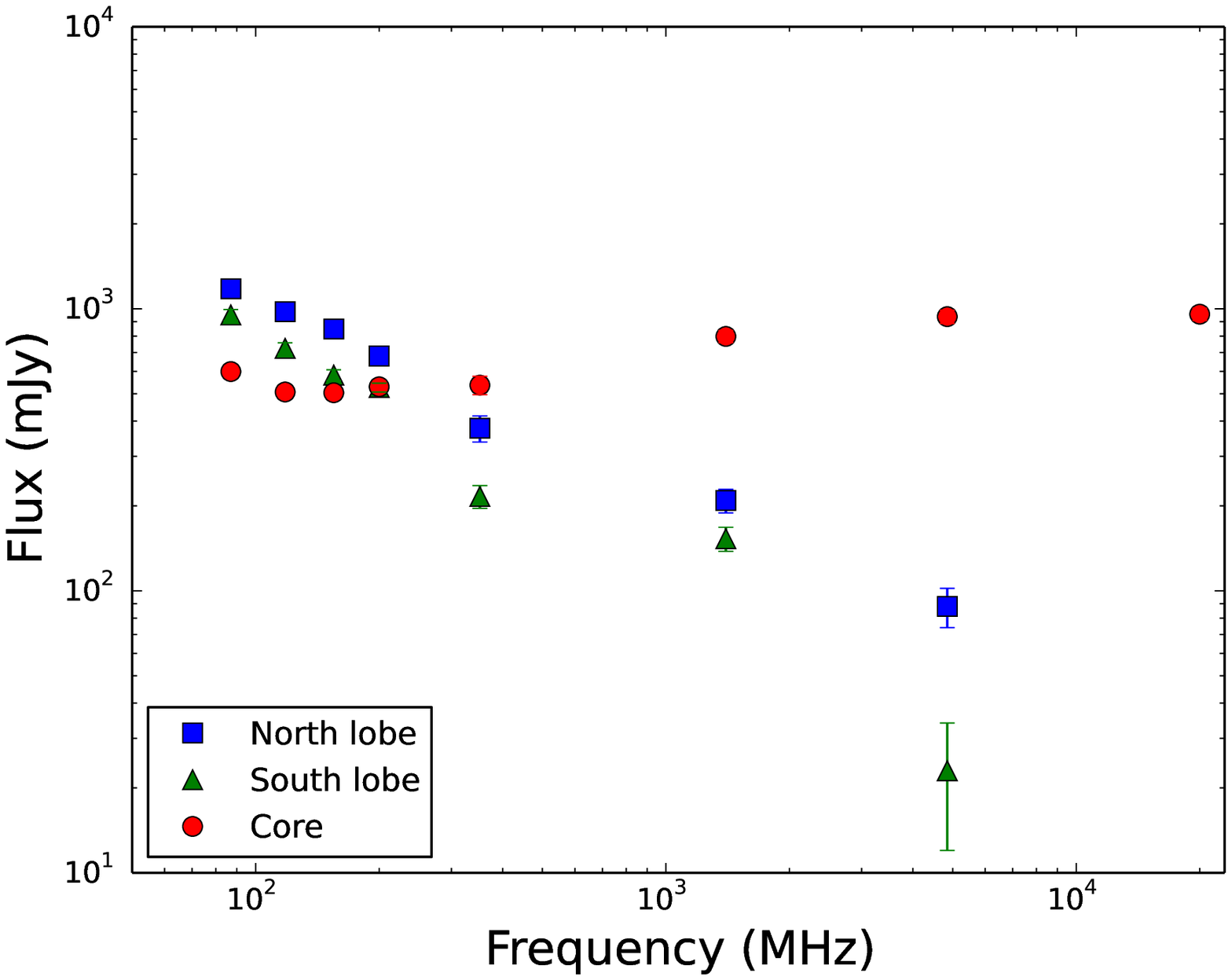}
\caption{(Left): NVSS image at 1.4 GHz of the field of \pks\ . Contours are traced at $(-1,1,2,4,8,16,32)\times1.7$ mJy beam$^{-1}$. The peak intensity is located at the core position and it is 747 mJy beam$^{-1}$. The restoring beam is circular with a 45 arcsec full width at half maximum (FWHM). (Right): Spectra of the north lobe (blue squares), core (red circles), and south lobe (green triangles). Additionally to NVSS data, we included the data up to 200 MHz from GLEAM \citep{gleam}; at 352 MHz from WISH \citep{wish}; the one at 5 GHz from the PMN survey \citep{pmn}. For the core, we added data from The Australia Telescope 20 GHz (AT20G) Survey \citep{atca}. \label{nvss}}
\end{figure*}

\section{\label{discusion}Discussion}

The classification of the nucleus of PBC\,J2333.9-2343 has been controversial in the literature, being classified as a Seyfert 2 in the optical, as unobscured AGN at X-rays, and as a blazar at radio frequencies \citep[see][]{parisi2012}. Since these classifications were obtained using data at different epochs, we obtained simultaneous coverage at X-ray, UV, optical, and radio frequencies, and found that these conflicting classifications cannot be related to variability issues\footnote{Note that the only difference is that we classified it as a type 1.9 AGN in the optical.}; instead they have to be attributed to an intrinsic property of the nucleus. Our observational results show that, in this source, both the jet and the nucleus are visible, with the jet closely pointing towards us, as we still have a direct view onto the BLR. This suggests that \pks\  is a blazar, explaining well all the classifications given at different wavebands.

The SED and radio analysis of this source suggest that the jet is observed at small angles. 
This line of sight is in good agreement with the lack of $N_{\rm H}$ inferred from the analysis of the X--ray data, which indicates that the source emission is not absorbed, as well as with the presence of a broad component in the optical spectrum, which suggests that the BLR is at least partly observed.
For this small $\theta_v$, and using public data from the NRAO VLA Sky Survey \citep[NVSS,][]{condon1998}, the apparent size of the large scale structure of \pks\ ($\sim22\arcmin$, or $ \sim 1150$ kpc, see left panel in Fig. \ref{nvss}) must correspond to an intrinsic deprojected value of $\sim 7$ Mpc for $\theta_v<10^\circ$, and to $> 13$ Mpc for $\theta_v<5^\circ$, as suggested by the SED results. The projected linear size of the largest giant radio galaxy known up to now is 4.69 Mpc \citep{machalski2008}.
Moreover, \cite{andernach2015} used a sample of 193 radio galaxies and determined that only one is a candidate to have a size larger than 5 Mpc. Thus, although it is highly implausible that the size of this source is so large, we leave open this possibility. Alternatively, this result might indicate that the direction of the jet has changed. This behaviour has been reported in the literature before, with changes in the direction of the jet as large as $\sim$90$^\circ$ \citep{giovannini2005}, and towards smaller angles too \citep[][]{evans1999,saripalli2002,konar2006,saikia2009, gopal2012, marecki2012,saripalli2013}. The case of \pks\  is exceptional, as the realignment of the jet made this galaxy pass from being a radio galaxy to become a blazar. Moreover, if we compare the p.a. of the jets in the \emph{VLA} and \emph{VLBA} data, with values of 139$^\circ$ and 133$^\circ$, respectively, it seems that this change has occurred in a direction that, by chance, conserved the projected p.a.

A change in the direction of the jet could be ascribed to restarting activity in the nuclei of the galaxies, i.e., when the nuclear activity at some point decreases or stops and then it restarts \citep{christiansen1973}. Observational indications of restarting activity can be inferred in radio galaxies from their extended emission. The old lobes can survive up to $\sim 10^7 - 10^8$ yr and trace the largest scale emission from the nucleus \citep{konar2006}, whereas the new activity can be observed at smaller scales. The result is that the different phases of activity can be observed in the same image. The analysis of the stellar populations shows that \pks\  is composed by stars on average older than $10^9$ years, therefore the galaxy has had enough time to form the old lobes and being subject to recurrent activity episodes.

The most striking evidence of restarting activity is given by the unambiguous morphology of DDRG and XRG. While we do not see indications of such morphologies in our data, the fact that the jet reorientation made this source to become a blazar explains well why we do not see the new lobes. Another indication in favour of this picture comes from the analysis of the radio spectra of the lobes versus that of the core. We have used data from the GLEAM \citep{gleam}, WISH \citep{wish}, NVSS \citep{condon1998}, PMN \citep{pmn}, and AT20G \citep{atca} surveys to estimate the flux densities of the core and the lobes at different frequencies (see Fig. \ref{nvss}, right panel). Here it can be seen that the emission of both lobes is decaying with a steep spectrum (-0.64 for the northern lobe, and -0.86 for the southern one), whereas the emission from the core is inverted (with a spectral index of 0.15), indicative of the presence of a compact core. While this might indicate ageing due to radiative losses in the lobes, the current data are not accurate enough to support such a claim because we cannot constrain a break at high frequencies.
Notwithstanding, \cite{giovannini2001} reported a correlation between the power of the core at 5 GHz, $P_c$, and the total power (lobes plus core) at 408 MHz, $P_T$, suggesting that sources with different kiloparsec-scale morphology and radio power are similar on parsec scales. For \pks\ we estimate $logP_T$ = 24.73 (from extrapolation from the emission at 352 MHz), thus from the correlation in \cite{giovannini2001} it is expected a value of $logP_C$(theoretical)=22.93. However, observationally we estimate a value of $logP_c$=24.65. This suggests either beaming (with $\theta < 20$ degrees), or restarting activity in \pks\ , as the nucleus is much brighter than expected. Although we cannot differentiate between these possibilities with the current data, it is worth noting that both are in agreement with a small viewing angle for the core component, and thus with a change in the direction of the jet.

Recurrent activity is hardly seen in radio galaxies that are not giant, as the conditions inside the cocoon might favour the propagation of the lobes only after a long time from the activity burst \citep{marecki2006}. Moreover, the selection of sources with indication of restarting activity is usually based on the radio morphology of the galaxies as DDRG or XRG \citep[see e.g.,][]{saikia2009}. Therefore we could be missing a fraction of restarted galaxies for orientation reasons, such as the case of \pks. Indeed, it has been suggested that DDRG could be XRG with a close-to-zero misalignment. \pks\  represents an extreme case of recurrent activity as all the following situations are occurring in the same source: (i) the jet has changed its orientation, (ii) the change has occurred on the plane of the sky, (iii) the jet is now pointing towards the observer, and (iv) the jet is relatively compact and therefore young, i.e. we do not find any evidence for a diffuse morphology around the core. 

Many studies have argued that the most plausible reason for the restarting activity in AGN is related to phenomenon of galaxy merging \citep{lara1999, schoenmakers2000,liu2004,marecki2006}, since this is able to trigger the nuclear activity \citep[e.g.,][]{stockton1983}. A visual inspection of the optical images in NED\footnote{http://ned.ipac.caltech.edu/} shows that there is a source located about 1' from \pks\ (RA=23h33m58.8s, DEC=-23d42m54s). This galaxy has a redshift of 0.05964 as estimated using an optical spectrum by \cite{jones2009}. 
Using these redshifts, we obtain that the luminosity distances are 193.3 Mpc for \pks\  and 239.3 Mpc for the neighbour, with a comoving distance of $\sim 35 h^{-1} Mpc$ (in case of a flat universe), indicating that the two objects are indeed quite far apart. 
Moreover, the difference in velocity between the two galaxies is  $\sim 3600$ $ km/sec$, which means that they can hardly be interacting. We recall that typical velocities in galaxy clusters are usually smaller than $1000 $ $km/s$ \citep[e.g.,][]{tempel2014}, although in some cases larger values have been reported \citep[e.g., 1900 $km/s$ for the AC\,114 cluster,][]{proust2015}.
Furthermore, recent works found that AGN triggering is strongly luminosity dependent \citep{treister2012}. Major mergers showed not to be the dominant AGN triggering process \citep[e.g.,][]{cisternas2011, schawinski2012,kocevski2012, villforth2014, villforth2017}, except for the most luminous AGN \citep[$\sim$ 10\%,][]{treister2012}, which is not the case of PBC\,J2333.9-2343. 
In contrast, it has been shown that most AGN (90\%) were triggered by secular processes such as minor interactions \citep[e.g.,][]{treister2012}.  In particular for radio galaxies, modest mergers have been proposed to explain the triggering mechanism of FRII galaxies, whereas hot accretion from the intra cluster medium or the interstellar medium has been proposed to explain the triggering mechanism of FRI galaxies \citep{tadhunter2016}. PBC\,J2333.9-2343 is intermediate between the FRI/FRII classification, therefore we cannot discard any of these possibilities as the triggering mechanism.

In agreement with the merging hypothesis, in a few examples of restarting activity galaxies, indications of binary SMBHs have been found \citep[see][and references therein]{gopal2012}. Recently, this has been suggested for example for J1328+2752 by \cite{nandi2016}. In their case, the optical spectrum of the source showed that the central component has double-peaked line profiles with different emission strengths. However, the resolution of our spectra is six times lower that the one of the SDSS spectra, therefore we cannot reject the possibility of unresolved double-peaked line profiles.

As mentioned above, we should also consider the possibility that the change in the direction of the jet could be related to changes/instabilities in the accretion disc. This occurrence has been formulated in terms of chaotic accretion for the SMBH feeding, which takes about $10^5$ years for the SMBH to grow \citep{king2015,schawinski2015}. While this timescale is compatible with the restarting activity timescale theoretically estimated by \cite{reynolds1997}, large amounts of mass would be required in order to change the spin axis of the SMBH and the direction of the jet \citep{nixon2013}. Although such large amounts of mass are not expected from accretion processes, we leave open this possibility as the origin of the restarted activity in \pks .

\section{\label{conclusion}Conclusions}

We performed a multiwavelength analysis of the nucleus of the GRG \pks\ using simultaneous data at radio, optical, UV and X-ray frequencies. We find that its optical spectrum is of a type 1.9 AGN, that it is unabsorbed at X-rays, and we classify it as a blazar from the analysis of the radio data and SED fitting. We therefore propose that the source is at the present time a blazar, where we have a direct view both to the jet and to the nucleus.

Indications of a change in the direction of the jet are observed when comparing \emph{VLBA} and \emph{VLA} data. This change has occurred in the plane of the sky, giving place to the actual morphology of the galaxy, which changed from being a radio galaxy to become a blazar. A comparison of the radio spectral indices of the lobes and the nucleus also favours this scenario, as the emission of the lobes is decaying, against the inverted spectrum of the nucleus. This is one of the few reports of episodic activity in an AGN where the source has not a DDRG or XRG morphology but where the indications come from restarting activity arguments. Since the selection of restarting activity is usually based on radio morphologies of the sources, we notice that we might be missing a non-negligible fraction of these galaxies, as the case of \pks\ should not be exceptional. 
A comparison of the spectra of the lobes and the nuclei of a large sample of radio galaxies would be an optimal test to find more restarting activity galaxy candidates and to provide a more reliable estimate of their fraction.

A neighbour galaxy is located close to this source, but we do not see indications of recent merging events. Although we cannot rule out a possible merger, we leave open the possibility of chaotic accretion being responsible for the changes observed in this galaxy. The accretion properties and detailed analysis of the optical spectra of the source will be discussed in a forthcoming paper.

The unique sensitivity and angular resolution of the Square Kilometer Array (SKA) will allow the discovery of a large number of galaxies with recurrent activity,
even among the youngest and smaller systems, drawing a statistical meaningful distribution of quiescent and active phases, fundamental to constrain the physics of accretion and ejection in galaxy history.

\begin{acknowledgements}

We are specially grateful to Dr. Parisi for leading the observing proposals.
Thanks to Dr. J. Masegosa, Dr. I. M\'{a}rquez, Dr. C. Cicone, Dr. E. Huerta, Dr. G. Migliori, Dr. P. Gandhi, Dr. S. Lotti, Dr. F. Nicastro, Dr. I. Saviane, Dr. A. Wolter, E.F. Jim\'{e}nez Andrade, Dr. C. Tadhunter, Dr. A. King, and Dr. R. Morganti for helpful and pleasant discussions during this work. 
We thank the comments from the anonymous referee, that helped improving the quality of the manuscript. 
LHG and FP acknowledge the ASI/INAF agreement number 2013-023-R1. MG acknowledges funding by a PRIN-INAF 2014 grant. MP acknowledges financial support from the Spanish Ministry of Economy and Competitiveness
(MINECO) through projects AYA2013-42227-P and AYA2016-76682C3-1-P, and from the Ethiopian Space Science and Technology Institute (ESSTI) under the Ethiopian Ministry of Science and Technology (MoST). We thank the STARLIGHT team for making their code public. In this work we used IRAF, the Image Analysis and Reduction Facility
made available to the astronomical community by the National
Optical Astronomy Observatories, which are operated by the Association
of Universities for Research in Astronomy (AURA), Inc.,
under contract with the US National Science Foundation.
The National Radio Astronomy Observatory is a facility of the National Science Foundation operated under cooperative agreement by Associated Universities, Inc.
This research made use of data obtained from the
\emph{XMM-Newton} Data Archive provided by the \emph{XMM}-Newton
Science Archive (XSA). This research made use of the NASA/IPAC
extragalactic database (NED), which is operated by the Jet Propulsion
Laboratory under contract with the National Aeronautics and Space
Administration. 
This research has made use of data provided by the High Energy Astrophysics Science Archive Research Center (HEASARC), which is a service of the Astrophysics Science Division at NASA/GSFC and the High Energy Astrophysics Division of the Smithsonian Astrophysical Observatory.

\end{acknowledgements}

\bibliographystyle{aa}
\bibliography{000referencias}

\end{document}